\documentclass[useAMS,usegraphicx,usenatbib,referee]{mn2e}

\title[Maximum momentum at modified shocks]
{The maximum momentum of particles accelerated at cosmic ray modified shocks}

\author[P. Blasi, E. Amato and D. Caprioli]
{Pasquale Blasi$^{1}$\thanks{E-mail: blasi@arcetri.astro.it},
Elena Amato$^{1}$\thanks{E-mail:amato@arcetri.astro.it} and
Damiano Caprioli$^{2}$\thanks{E-mail:caprioli@sns.it}\\
$^{1}$INAF-Osservatorio Astrofisico di Arcetri, 
Largo E. Fermi, 5, 50125, Firenze, Italy\\
$^{2}$Scuola Normale Superiore, Pisa, Italy}

\begin{document}

\date{Accepted ----. Received -----}


\maketitle

\label{firstpage}

\begin{abstract}
Particle acceleration at non-relativistic shocks can be very
efficient, leading to the appearance of non-linear effects due to the
dynamical reaction of the accelerated particles on the shock structure 
and to the non-linear amplification of the magnetic field in the shock
vicinity. 
The value of the maximum momentum $p_{max}$ in these circumstances
cannot be estimated using the classical results obtained within
the framework of test particle approaches. We provide here the first
attempt at estimating $p_{max}$ in the cosmic ray modified regime,
taking into account the non-linear effects mentioned above. 
\end{abstract}

\begin{keywords}
acceleration of particles - shock waves
\end{keywords}

\section{Introduction}

Since the beginning of the investigations on the mechanism of particle 
acceleration at shock fronts, estimating the maximum energy of the
accelerated particles has been one of the main goals, in the
perspective of assessing the ability of the mechanism to explain the
observed energies in cosmic rays. This issue has played a
crucial role especially in the context of the so-called supernova 
paradigm for the origin of galactic cosmic rays. In fact, 
it was understood very soon, in the case of supernova 
remnants' shocks, that acceleration would not be efficient
enough and provide sufficiently high energies unless magnetic 
scattering could be self-generated by the particles themselves through 
their motion in the upstream plasma. The mechanism through which this
generation of waves could occur was identified in the streaming
instability (\cite{bell78}) and widely discussed later by
\cite{lc83a,lc83b}. In these latter papers, in particular, the maximum 
energy was estimated for shocks in supernova remnants, already
believed to be candidate sources of galactic cosmic rays. The result
of this investigation was however that the maximum energy, even in the
presence of self-generated scattering agents, is of the order of
$\sim 10^3-10^4$ GeV, quite lower than the energy at which the knee is
observed. Today we know that this energy is also lower than the
maximum energy observed in the proton component by the KASCADE
experiment (see \cite{kascade} for a comprehensive review of
experimental results in the knee region) that seems to be
as high as $\sim 10^7$ GeV (see \cite{blasirev} for a recent review).

More recently the problem has been complicated even more by the
understanding that particle acceleration can be rather efficient and
break the {\it test particle} regime that old calculations were based
upon. This important result was obtained independently within
completely different approaches, two-fluid models
(\cite{dr_v80,dr_v81}), kinetic models 
(\cite{malkov1,malkov2,blasi1,blasi2,amato1,amato2}) and numerical
approaches, using both Monte Carlo and other simulation procedures 
(\cite{je91,bell87,elli90,ebj95,ebj96,kj97,kj05,jones02,bykov}).
The non-linear dynamical reaction of the accelerated particles
introduces a positive reaction that results in a flattening of the
high energy spectra, which are no longer power laws, thereby
enhancing the role of particles at the highest energies. Moreover a
gradient in the velocity of the upstream fluid is produced by the
pressure of the accelerated particles. The value of $p_{max}$ is
important to determine the spectrum, and the spectrum determines the
amplification of the magnetic field which in turn determines
$p_{max}$. The formalism of \cite{lc83a,lc83b} is not appropriate to
describe this type of situation.  

In this paper we describe the first calculation of the maximum
momentum for cosmic ray modified shocks, including all the relevant 
non-linear effects. The description of the acceleration process, with
the dynamical reaction of cosmic rays and the magnetic field
amplification taken into account, is based on the recent papers by
\cite{amato1,amato2}.  

The paper is organized as follows: in \S \ref{sec:method} we describe
the method that allows us to compute the acceleration time as a
function of the particle momentum; in \S \ref{sec:pmax} we describe
the procedure used to determine $p_{max}$ in the case of modified 
shocks and discuss the underlying assumptions; in \S \ref{sec:results}
we present the results we find for the maximum achievable momentum
in supernova remnant shocks; finally a critical discussion of the
method adopted and a summary of the main results is provided in \S~
\ref{sec:critical}.

\section{Acceleration time for modified shocks} 
\label{sec:method}

The full description of the calculations that allow us to compute
the spectrum of the accelerated particles, the dynamics of the
upstream and downstream plasmas and the strength of the amplified
magnetic field due to cosmic rays' streaming instability in the 
stationary regime can be found in the papers by
\cite{amato1,amato2}. 

In this section we present the main achievement of our work, which
consists of a derivation of the acceleration time $t_{acc}(p)$ (as a
function of the momentum $p$) for particles accelerated at a shock
with an arbitrary level of shock modification. The expression we
derive is a generalization of the well known formulae found in the 
context of quasi-linear theory (\cite{lc83a,lc83b}) and reduces to them
when the non-linear shock modification is negligible. 

In general, the transport of the accelerated particles is described 
by the time dependent diffusion-convection equation:
\begin{equation}
\frac{\partial f}{\partial t} + u \frac{\partial f}{\partial x} = 
\frac{\partial}{\partial x}\left[ D \frac{\partial f}{\partial x}
  \right] + \frac{1}{3} \frac{du}{dx} p\frac{\partial f}{\partial p} +
Q(x,p), 
\label{eq:transport}
\end{equation} 
where $f(t,x,p)$ is the distribution function of the accelerated
particles, $D(t,x,p)$ is the spatial diffusion coefficient and $Q$ is
the injection term, taken in the form $Q(x,p)=Q_0(p) \delta(x)$, with 
the assumption that injection takes place only at the shock surface 
on the downstream side (although this assumption is not strictly
needed). The $x-$axis goes from upstream
infinity ($x=-\infty$) to downstream infinity ($x=+\infty$) and the
shock is at $x=0$. We will use the index $1$ ($2$) to refer to the 
region immediately upstream (downstream) of the shock. 

The problem of determining the time scale associated with the change
of the distribution function around the maximum momentum is impossible
to face analytically in all its complexity, as implied by the
non-linearity of the system: in general, $f$, $u$ and $D$ all depend
on time, and change on comparable time scales. Moreover, $f$ depends in
a complicated way on all other quantities (\cite{amato1,amato2}). 

Let us assume that we know an approximation of the solution $f$ in 
the form of a solution of the stationary problem for a given value 
of the maximum momentum $p_{max}$, as can be calculated following
\cite{amato1,amato2}. This value of $p_{max}$ can be considered as  
a {\it guess} value to start with. In this case the velocity profile
$u(x)$ can be considered as determined in first instance from the
guess function $f$ as derived by the assumption of quasi-stationarity,
using the calculations of \cite{amato1,amato2}. Within this
approximate framework, we can introduce the Laplace transform of the
distribution function with respect to time:
\begin{equation}
g(s,x,p) = \int_0^\infty dt\ e^{-st} f(t,x,p), 
\end{equation}
so that the transport equation becomes: 
\begin{equation}
s g + u \frac{\partial g}{\partial x} = 
\frac{\partial}{\partial x}\left[ D \frac{\partial g}{\partial x}
  \right] + \frac{1}{3} \frac{du}{dx} p\frac{\partial g}{\partial p} +
\frac{1}{s} Q(x,p), 
\label{eq:transportLaplace}
\end{equation}

Integration of Eq.~\ref{eq:transportLaplace} in the vicinity of the shock
(namely between $x=0^-$ and $x=0^+$) leads to the following boundary
condition at the shock:
\begin{equation}
\left[D\frac{\partial g}{\partial x}\right]_2 -
\left[D\frac{\partial g}{\partial x}\right]_1 
+ \frac{1}{3} (u_2-u_1) p \frac{\partial g_0(s,p)}{\partial p} + 
\frac{1}{s} Q_0(p) = 0,
\label{eq:jump}
\end{equation}
where $u_1=u(0^+)$, $g_0(s,p)=g(s,x=0,p)$ and we have used the 
continuity of the distribution function (at any time) at the shock.  

We can find another condition on the distribution function by
integrating the transport equation (Eq.~\ref{eq:transportLaplace}) 
in the upstream section from $x=-\infty$ to $x=0^-$, so as to obtain:
\begin{equation}
s I(s,p) + u_1 g_0(s,p) -
\int_{-\infty}^{0^-} dx\ g(s,x,p) \frac{du(x)}{dx} = \left[ D
  \frac{\partial g}{\partial x}\right]_1 + \frac{1}{3}
p\frac{\partial}{\partial p} \int_{-\infty}^{0^-} dx\ g(s,x,p)
\frac{du(x)}{dx},
\label{eq:jump2}
\end{equation}
where $I(s,p) = \int_{-\infty}^{0^-} dx\ g(s,x,p)$. Following 
\cite{blasi1} we introduce the quantity 
\begin{equation}
u_p(s,p) = u_1 - \frac{1}{g_0} \int_{-\infty}^{0^-} dx\ g(s,x,p)
\frac{du}{dx},
\label{eq:up}  
\end{equation}
which allows us to rewrite Eq.~\ref{eq:jump2} as
\begin{equation}
\left[ D\frac{\partial g}{\partial x}\right]_1 =
s I(s,p) + g_0 u_p \left( 
1 + \frac{1}{3} \frac{d\ln u_p}{d \ln p}
\right) - \frac{1}{3} p \frac{\partial g_0}{\partial p} (u_1 - u_p).
\label{eq:D1}
\end{equation}
Clearly in this equation $u_p$ depends on time through the functions
$f$ and $u$, which again reflects the non-linearity of the underlying 
equation.

The transport equation in the downstream fluid can be considerably
simplified if we assume that the fluid velocity is independent of
location $u(x)=u_2$ and that the diffusion coefficient is also
independent of $x$. In this case, for any $x\neq 0$ Eq.~\ref{eq:transport}
becomes:
\begin{equation}
\frac{\partial f}{\partial t} + u_2 \frac{\partial f}{\partial x} = 
D_2 \frac{\partial^2 f}{\partial x^2}.
\label{eq:downtransport}
\end{equation}
If we use again the Laplace transform of the distribution function
with respect to time we have
\begin{equation}
s g + u_2 \frac{\partial g}{\partial x} = D_2 \frac{\partial^2
  g}{\partial x^2},
\label{eq:downlaplace}
\end{equation}
whose solution has the form $g(s,x,p) = g_0(p) e^{-\beta x}$ (in the
downstream region $x>0$) with $\beta$ easily found by substitution, so 
that:
\begin{equation}
g(s,x,p) = g_0(p) \exp\left\{ \frac{u_2 x}{2 D_2}
\left[1-\left(1+\frac{4sD_2}{u_2^2}\right)^{1/2}\right]\right\}.
\label{eq:downsol} 
\end{equation}
This implies that 
\begin{equation}
\left[ D\frac{\partial g}{\partial x}\right]_2 = 
\frac{1}{2} g_0(p) u_2 \sigma_2, ~~~~~~~~
\sigma_2 = 1 - \left(1+\frac{4sD_2}{u_2^2}\right)^{1/2}. 
\label{eq:D2}
\end{equation}
At this point, substituting Eqs.~\ref{eq:D1}-\ref{eq:D2} into the jump
condition Eq.~\ref{eq:jump} we obtain the following equation:
\begin{equation}
g_0 \left\{ \frac{1}{2}u_2 \sigma_2 - u_p \left(1+\frac{1}{3}\frac{d\ln
  u_p}{d \ln p}\right)\right\} + \frac{1}{3} (u_2-u_p) p
  \frac{dg_0}{dp} -s I +\frac{1}{s} Q_0 = 0.
\label{eq:g0}
\end{equation}
As already proposed by \cite{malkov1} and \cite{amato1}, it is useful
to rewrite the transport equation in an implicit form by integrating
it between $x=-\infty$ and a generic point $x$ upstream. We repeat
this procedure here on the Laplace-transformed transport equation 
(Eq.~\ref{eq:transportLaplace}).
This leads to the following implicit expression for $g$:
$$
g(s,x,p) = \exp\left\{ \int_0^x dx' \frac{u(x')}{D(x',p)} \right\}
\left\{ g_0(s,p) + \int_0^x dx' \frac{1}{D(x',p)} 
\exp\left[ - \int_0^{x'} dx'' \frac{u(x'')}{D(x'',p)} \right]\times
\right.
$$
\begin{equation}
\left.
\left\{ -\int_{-\infty}^{x'} dx'' g(s,x'',p) \frac{du}{dx''}
\left(1+\frac{1}{3}\frac{d\ln g(s,x'',p)}{d\ln p}\right)
+ s \int_{-\infty}^{x'} dx''
g(s,x'',p)\right\}\right\}.
\end{equation}
As already discussed by \cite{malkov1} and \cite{amato1}, 
one can approximate this complex implicit solution as 
$$
g(s,x,p) = g_0 (s,p) \exp\left\{ \frac{q(p)}{3}
\int_0^x dx' \frac{u(x')}{D(x',p)}\right\} +
$$
\begin{equation}
~~~~~~~~~~~~~~~~~~~~~~~~~~~
+ s \exp\left[\int_0^x dx' \frac{u(x')}{D(x',p)} \right] 
\int_0^x dx'\frac{1}{D(x',p)} \int_{-\infty}^{x'} dx'' g(x'') 
\label{eq:approx}
\end{equation}
where $q(p)$ is now the slope of the distribution function at the
shock. This expression is obtained by expanding the previous
expression for small values of $\int_0^x dx'
\frac{u(x')}{D(x')}$. In \cite{amato1} it was shown {\it a posteriori}
that this is a good approximation of the spectral distribution of the
particles as a function of the spatial coordinate.

The agreement was found to be worse when the shock
is quasi-linear, namely for small modifications. The reason for this
is in the fact that Eq.~\ref{eq:approx} satisfies the jump condition
in the stationary regime ($s\to 0$) only when $u_2\to 0$. A better
approximation for $g$ is in fact of the form, 
$$
g(s,x,p) = g_0 (s,p) \exp\left\{ \frac{q(p)}{3}
\left(1-\frac{u_2}{u_1}\right)
\int_0^x dx' \frac{u(x')}{D(x',p)}\right\} +
$$
\begin{equation}
~~~~~~~~~~~~~~~~~~~~~~~~~~~
+ s \exp\left[\int_0^x dx' \frac{u(x')}{D(x',p)} \right] 
\int_0^x dx'\frac{1}{D(x',p)} \int_{-\infty}^{x'} dx'' g(x'') 
\label{eq:approx1}
\end{equation}
which satisfies the jump condition exactly. Note that
Eq.~\ref{eq:approx1} differs from Eq.~\ref{eq:approx} only in that
at given momentum the particle distribution is slightly more 
extended spatially. In the following we adopt Eq.~\ref{eq:approx1} as an 
approximation of the spatial distribution of the accelerated
particles. In evaluating the term $s I(s,p)$ in Eq.~\ref{eq:g0} we
retain only the first term of $g$ in Eq.~\ref{eq:approx1}, since in
the end we will be interested in the limit $s\to 0$ (large times). 
This means that Eq.~\ref{eq:g0} becomes:
\begin{equation}
g_0 \left\{ \frac{1}{2}u_2 \sigma_2 - s \Lambda(p) 
- u_p \left(1+\frac{1}{3}\frac{d\ln
  u_p}{d \ln p}\right)\right\} + \frac{1}{3} (u_2-u_p) p
  \frac{dg_0}{dp} +\frac{1}{s} Q_0 = 0,
\label{eq:g00}
\end{equation}
where
\begin{equation} 
\Lambda(p)=\int_{-\infty}^0 dx \exp\left\{ \frac{q(p)}{3}
\left(1-\frac{u_2}{u_1}\right)
\int_0^x dx' \frac{u(x')}{D(x')}\right\}.
\label{eq:lambda}
\end{equation}
The solution of this equation is easily found to be:
$$
g_0(s,p) = \frac{1}{s} \frac{\eta n_{gas,1}}{4\pi p_{inj}^3}
\frac{3R_{sub}}{R_{tot}U_p(p)-1} \exp \left\{ -\int_{p_{inj}}^p
\frac{dp'}{p'} \frac{3U_p(p') R_{tot}}{R_{tot} U_p(p') -1}\right\}\times
$$
\begin{equation}
\times \exp \left\{ \int_{p_{inj}}^p
\frac{dp'}{p'} \frac{3}{2} \frac{\sigma_2}{R_{tot} U_p(p') -1}\right\}
\times \exp \left\{ -\int_{p_{inj}}^p 
\frac{dp'}{p'} \frac{3 s \Lambda(p')}{R_{tot} U_p(p') -1}\right\},
\end{equation}
or, in a more compact form, 
\begin{equation}
g_0(s,p) = \frac{1}{s} K(p) \exp\left[h(p,s)\right],
\label{eq:compact}
\end{equation}
where
\begin{equation}
K(p) = \frac{\eta n_{gas,1}}{4\pi p_{inj}^3}
\frac{3R_{sub}}{R_{tot}U_p(p)-1} \exp \left\{ -\int_{p_{inj}}^p
\frac{dp'}{p'} \frac{3U_p(p') R_{tot}}{R_{tot} U_p(p') -1}\right\}
\end{equation}
and
\begin{equation}
h(p,s) = \int_{p_{inj}}^p \frac{dp'}{p'} \left[
\frac{3}{2} \frac{\sigma_2}{R_{tot} U_p(p') -1} - \frac{3 R_{tot}}{u_0} 
\frac{s \Lambda(p')}{R_{tot} U_p(p') -1} 
\right],
\end{equation}
with $U_p(p)=u_p/u_0$ and $u_0=u(-\infty)$.

At this point we follow the elegant procedure discussed by
\cite{drury83}. We need to calculate the inverse Laplace transform
$$
f(t,x,p) = \frac{1}{2\pi i} \int_{-i\infty}^{+i\infty} ds\ 
\frac{e^{ts}}{s}\ K(p)\ e^{h(p,s)}.
$$
We introduce the function 
\begin{equation}
\phi(p,t) = \frac{1}{2\pi i} \int_{-i\infty}^{+i\infty} ds ~ e^{ts} 
e^{h(p,s)}.
\end{equation}
One can easily show that the inverse Laplace transform of the function 
\begin{equation}
\tilde f (p,t)= K(p) \int_0^t dt'\ \phi(p,t')
\end{equation}
is exactly the function $g_0$ as written in Eq.~\ref{eq:compact}. It
follows that the solution of our problem is in fact $f=\tilde f$. 
In the approximate framework in which we are, the function $\phi$, as
in the linear case discussed by \cite{drury83}, represents the
probability that a particle starting with energy $p_{inj}$ has reached
the momentum $p$ in a time $t$ ($\phi$ is normalized to unity). The
mean time for the acceleration to momentum $p$ is therefore
$<t(p)> = \int_0^\infty dt\ t\ \phi(p,t)$. One can use the following obvious
property of the function $\phi$ 
\begin{equation}
\int_0^\infty dt\ \phi(p,t)\ e^{-ts} = \exp\left[h(p,s)\right],
\label{eq:prop}
\end{equation}
to determine $<t>$, by deriving Eq.~\ref{eq:prop} with respect to $s$
and calculating it at $s=0$. From this follows that:
\begin{equation}
<t> = - \left[\frac{\partial h}{\partial s}\right]_{s=0} = 
\frac{3 R_{tot}}{u_0^2}\
\int_{p_{inj}}^p \frac{dp'}{p'} \left\{
  \frac{R_{tot} D_2(p')}{R_{tot}U_p(p')-1} + \frac{u_0 \Lambda(p')}
{R_{tot} U_p(p') -1} \right\}. 
\label{eq:tacc}
\end{equation}
For an unmodified shock, with spatially constant fluid velocities
upstream and downstream, one has that $q(p)=3u_1/(u_1-u_2)$ and the
previous formula easily reduces to the well known:
\begin{equation}
<t> = \frac{3}{u_1 - u_2} \left[ \frac{D_2}{u_2} + \frac{D_1}{u_1}
\right],
\label{eq:lagage}
\end{equation}
if the diffusion coefficient is in the form of Bohm diffusion. For an
arbitrary form of the diffusion coefficient the expression in 
Eq.~\ref{eq:lagage} is correct within a factor of order unity. 

\section{Calculation of $p_{max}$}
\label{sec:pmax}

In general the value of $p_{max}$ is determined by the competition
between the acceleration time and the shortest among the finite age of
the source, the time scale for energy losses, and 
the time scale for escape of particles with momentum $p_{max}$ from
upstream infinity.  

Here we concentrate on the self-consistent calculation of the
maximum momentum in those cases when it is determined by the finite
age of the source (this is expected to be the case for supernova
remnants). In such a context each value of $p_{max}$ therefore
reflects a different age $\tau$ of the system. In this sense, our
previous work (\cite{amato1,amato2}) was based on the implicit
assumption that a quasi-stationary regime could be reached on time
scales shorter than $\tau$, but no specific recipe for the calculation
of $p_{max}$ was adopted. In other work on non-linear diffusive
particle acceleration at shock waves, $p_{max}$ was either fixed or
determined according with a simple recipe based on test-particle
theory. 

Our determination of the maximum momentum proceeds through 2 different
steps: 1) we first determine the acceleration time $t_{acc}(p)$
(Eq. \ref{eq:tacc}) for a given value of $p_{max}^{(k)}$ (at the 
first step this is just a guess value $p_{max}^{(0)}$); 2) we compare 
$t_{acc}(p_{max}^{(k)})$ with the age of the system $\tau$ and
determine an updated value of $p_{max}$, say $p_{max}^{(k+1)}$. 
The procedure is iterated until $t_{acc}(p_{max})=\tau$, where 
$p_{max}$ is now the actual maximum momentum for our problem.
It is crucial to keep in mind that due to the non-linearity of the
system, changing $p_{max}$ causes the all system to change (velocity
profile, spectrum of accelerated particles, efficiency of particle 
acceleration and also the diffusion coefficient, if the scattering is
due to self-generated waves). 

The results obtained by applying this procedure are illustrated in the
next section. 

\section{Results}
\label{sec:results}
\begin{figure}
\resizebox{\hsize}{!}{
\includegraphics{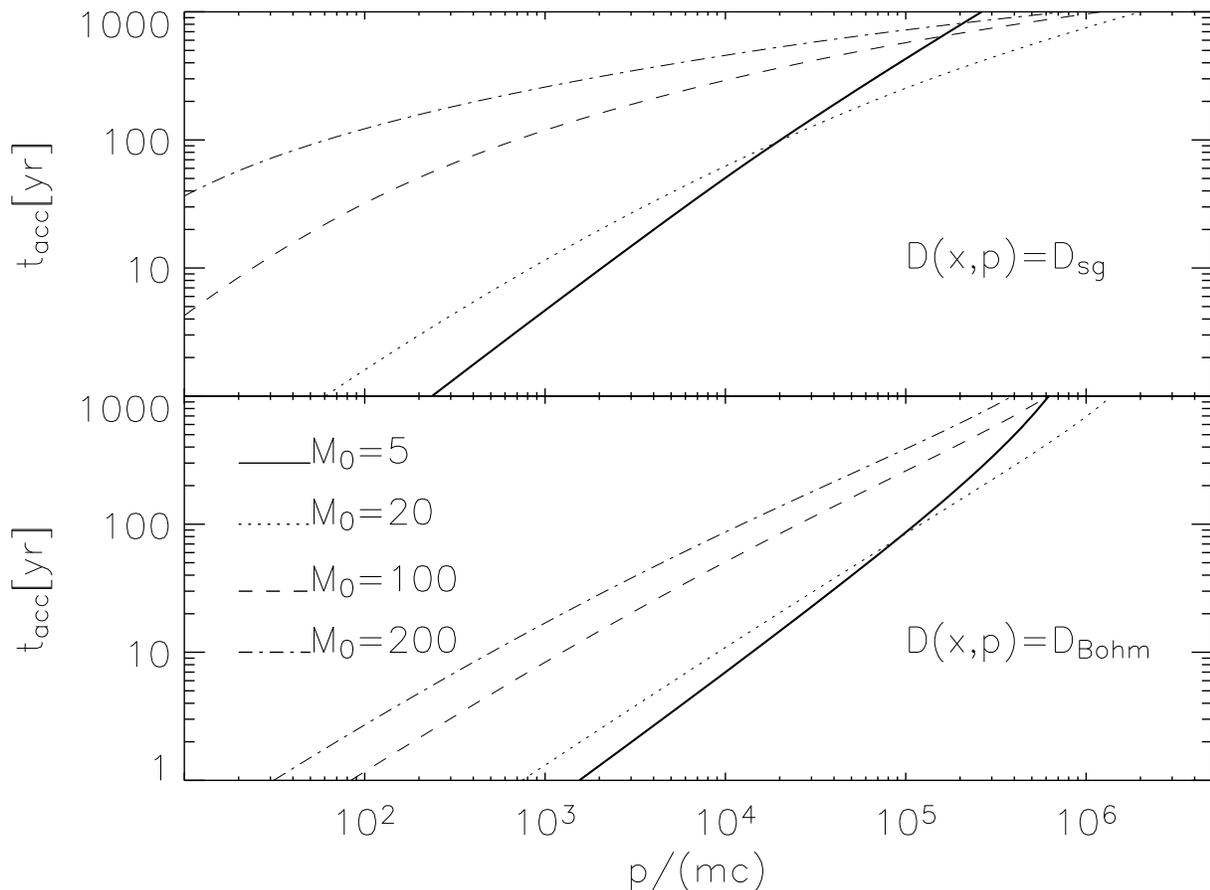}
}
\caption{We plot the time (in years) it takes for the system to
  accelerate particles up to a momentum $p$ (in units of $mc$) under
  different conditions. The different panels refer to different
  choices of the diffusion coefficient. In the top panel the diffusion
  coefficient is the self-generated one, while in the bottom panel 
  we assume Bohm diffusion in the amplified magnetic field (see text for
  details). The different line types refer to different values of the
  shock Mach number: $M_0=5$ for the solid line, $M_0=20$ for the dotted line,
  $M_0=100$ dashed, $M_0=200$ for the dot-dashed line.}
\label{fig:tacc}
\end{figure}

Following the method outlined in \S~\ref{sec:method} we computed the 
acceleration time as a function of the particle momentum $p$ for a
system with $\tau=1000\ yr$, under different assumptions on the
shock Mach number at infinity $M_0$ and on the diffusion properties of
the system. The results of these calculations are summarized in 
Fig.~\ref{fig:tacc}. We assumed a velocity of the shock wave of 
$5\times 10^3 km/s$ and considered values of $M_0$ ranging from $5$
(unmodified shocks) to $200$ (already in the asymptotic regime for 
strongly modified shocks). The injection
momentum was taken to be $\xi\ p_{therm}$ with $p_{therm}$ corresponding
to the maxwellian temperature $T_2$ and $\xi=3.5$. We also related the 
injection efficiency $\eta$ to the compression ratio at the subshock
through the recipe proposed by \cite{vannoni} and already adopted 
by \cite{amato1,amato2}. In the top panel of
Fig.~\ref{fig:tacc} we plot the acceleration time up to a momentum
$p$ for $D(x,p)=D_{sg}$, namely with self-generated scattering, with 
$D_{sg}=(4/3\pi) v r_L/\mathcal{F}$ the result of quasi-linear theory 
for scattering induced by self-generated Alfv\'en waves. Here $v$ and 
$r_L$ are the particle velocity and Larmor radius (in the background 
field $B_0$) respectively, and $\mathcal{F}$ is the dimensionless wave 
energy density per unit logarithmic band-width (see e.g. \cite{amato2}
for details). In the bottom panel of Fig.~\ref{fig:tacc} we show the
acceleration time for $D(x,p)=D_{Bohm}$, with $D_{Bohm}$ referring to
Bohm-like diffusion in the amplified magnetic field $\delta B$ found by 
integrating $B_0^2\mathcal{F}$ in $k$-space: $D_{Bohm}=(1/3) v r_L 
(B_0/\delta B)$. In all of our calculations the background magnetic
field was taken to be $B_0=10 \mu G$. 

The maximum momentum obtained from our
iterative procedure is seen as the intersection with the horizontal
line $\tau=1000$ years, assumed to be the age of the system. 

Each of the curves in Fig.~\ref{fig:tacc} was obtained by following
the procedure outlined in \S~\ref{sec:pmax}.The first thing to notice 
by comparing the two different panels in Fig.~\ref{fig:tacc} is that
the maximum achievable momentum, for a given Mach number, is roughly
the same (within a factor of a few (2-4) at most) for the two
diffusion models adopted. This observation is very important given the
fact that the diffusion coefficient is basically unknown and not
easily obtained from observations.

Still the comparison between the two panels allows 
to highlight what the main effect of the different diffusion models is:
in the case of strongly modified shocks the spectra become very flat at 
high energies as discussed in detail elsewhere (e.g. \cite{amato2}) and
most of the power of the Alfv\'en waves excited by the streaming cosmic rays
is deposited in long wavelength waves (the ones that are resonant with
high energy particles), with a relatively smaller fraction going into waves 
that are able to efficiently scatter the low energy particles. This is the
reason for the behaviour of $t_{acc}(p)$ observed in the upper panel,
which may seem surprising at first sight for the flatness of the curves
referring to $M_0=100-200$. By comparing the two panels, it is clear that
things change drastically, with the dependence of  $t_{acc}$ on $p$ becoming
much stronger, if the energy of the waves is redistributed 
uniformly between different wavelengths, which is what one is effectively
assuming by adopting a Bohm-like diffusion coefficient in the amplified 
magnetic field.
    
\begin{figure}
\resizebox{\hsize}{!}{
\includegraphics{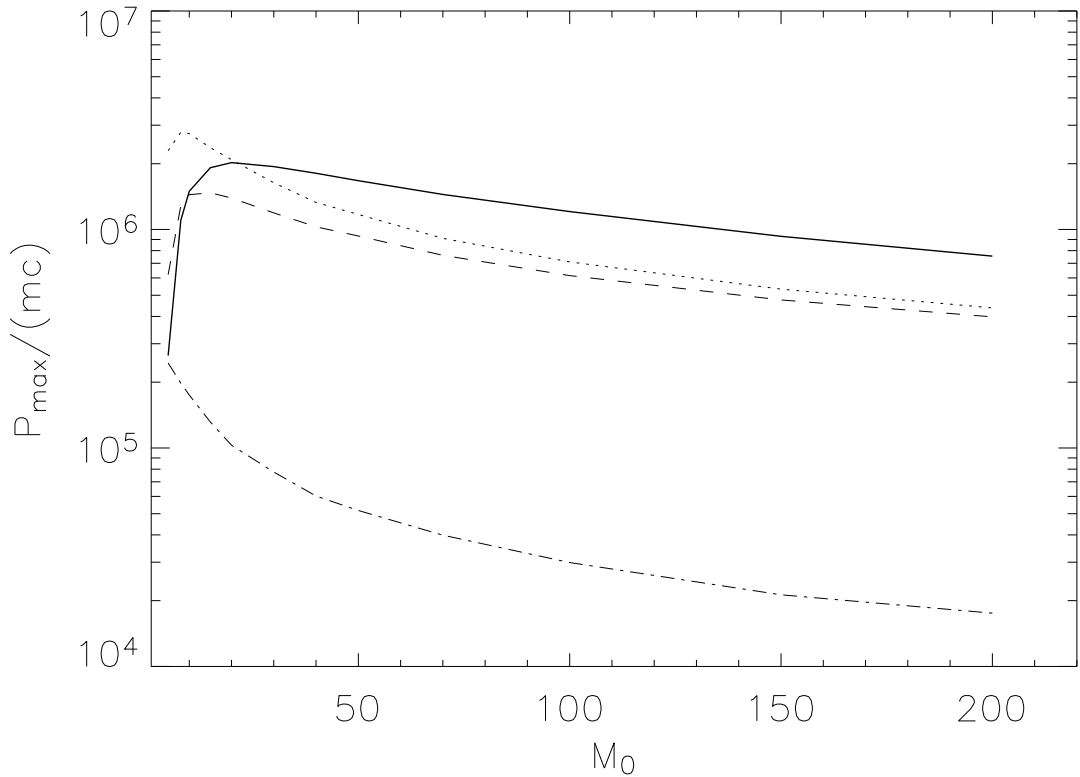}
}
\caption{We plot the maximum momentum the particles can achieve within
  a time $\tau=10^3\ {\rm yr}$ as a function of the Mach number of the
  shock. The different curves refer to different choices of the
  diffusion coefficient: the solid line is for self-generated
  diffusion; the dashed line is for Bohm diffusion in the
  self-amplified magnetic field; the dotted line is for Bohm diffusion
  in a magnetic field that is spatially constant in strength and equal
  to $\delta B(0)$, the value reached at the shock location by the
  cosmic ray streaming induced field; finally the dot-dashed line is 
  for Bohm diffusion in the background magnetic field.}
\label{fig:pmax}
\end{figure}

Finally, let us discuss the behaviour of the maximum achievable 
momentum as a function of the Mach number. This is qualitatively 
the same for both models of diffusion: $p_{max}$ initially increases 
quickly with the Mach number as a result of 
field amplification in modified shocks and then decreases somewhat, 
although slowly, when $M_0$ is further increased. This latter fact is
clearly related to the more strongly modified velocity profile that
leads to lower and lower values of the flow velocity in the vicinity 
of the shock and as a consequence increasingly longer residence
times upstream.
The dependence of the maximum achievable momentum on the shock Mach number 
is illustrated more clearly in Fig.~\ref{fig:pmax}, where we plot $p_{max}$
as a function of $M_0$ for four different models of diffusion. The plot
refers again to a system with $\tau=10^3\ {\rm yr}$ and all other values of 
the relevant parameters are the same as above, including the background 
magnetic field strength and the recipe for $p_{inj}$. 

The solid curve refers to $D(x,p)=D_{sg}(x,p)$ as described above, the 
dashed curve is for $D(x,p)=D_{Bohm}(x,p)$, the dotted curve is for 
$D(x,p)=D_{Bohm}(0,p)$ (namely the diffusion coefficient equals the
Bohm diffusion calculated in the amplified magnetic field at the shock
surface, without spatial variation) and finally the dot-dashed curve 
refers to Bohm diffusion in the background magnetic field. The first
aspect that clearly appears from the figure is the enhancement
of $p_{max}$ that arises from magnetic field amplification. For high Mach 
number shocks the magnetic field strength at the shock is amplified by a 
factor of order 25 and this implies an enhancement by a factor 10-30   
(depending on the diffusion model adopted) of the maximum achievable energy
(comparison of the solid, dashed or dotted curves with the dot-dashed one).
On the other hand, $p_{max}$ does not strongly depend on the detailed 
diffusion model adopted, in terms of the turbulence spectrum, with the
difference between Bohm diffusion and resonant scattering with
self-generated Alfv\'en turbulence (comparison between the solid and
dashed curves) being a factor of 2 at most.  
A few remarks are worth being made. Note that the values of the
parameters adopted throughout our calculations are reminiscent of
those appropriate to the case of young supernova remnants (SNRs). This is 
because we wished to address here the issue of the maximum energy 
that can be achieved in this class of sources, as candidate sources 
of galactic cosmic rays. Moreover there seems to be now solid evidence
for amplified magnetic fields in SNRs (\cite{warren}), so that there is
observational support for the idea that cosmic rays might be
efficiently accelerated at SNRs' shocks. The shock modification is a
natural consequence of the effectiveness of these shocks as cosmic ray 
accelerators. 
 
The maximum energy that the accelerated particles can 
achieve in the context of modified shocks is always less than 
one would naively guess by using a \cite{lc83a} type of estimate with 
the upstream velocity set equal to $u_0$ and the magnetic field
evaluated at the shock surface. This is due to the combination of two
effects. On one side, the net effect of the shock modification is to 
slow down the upstream fluid (precursor), so that the effective
velocity {\it felt} by the particles of given momentum is always lower 
than $u_0$, thereby making the acceleration process slower. This is 
illustrated effectively through the dotted line in Fig. \ref{fig:pmax}, 
where the diffusion coefficient is taken as Bohm-like calculated with 
the magnetic field at the shock, assumed constant in the precursor. In
this case the value of $p_{max}$ is always lower than $\sim 10^6$
GeV/c for Mach number $M_0>100$, while in the absence of shock
modification one would get $p_{max}\sim 6\times 10^6$ GeV/c. This
factor $>5$ is solely due to the existence of a precursor. 
There is however a second (less important) effect that causes a 
further decrease of $p_{max}$: the spatial
distribution of the accelerated particles in the precursor also
determines a spatial profile of the strength of the magnetic field
associated with the self-generated Alfv\'en waves, so that the effective 
diffusion coefficient is also larger 
than $D_{Bohm}(0,p)$. This effect is visible when comparing the dashed
line with the dotted line in Fig. \ref{fig:pmax}. The combination of
the two effects leads to an overall decrease of $p_{max}$ by a factor
$\sim 10$ compared with the naive expectation {\it a la}
\cite{lc83a,lc83b}, more evident for higher values of the Mach number. 

\section{A critical discussion of the method and summary of the main
  results}
\label{sec:critical}

In this section we summarize the reasons for the need of the type of
calculations presented here and we discuss some shortcomings of the
proposed mathematical approach. 

To our knowledge this represents the first attempt to determine the
maximum momentum of the particles accelerated at a cosmic ray modified
shock, where the nonlinear dynamical reaction of the accelerated
particles changes the shocked fluid, the particles self-generate their
own scattering centers, and the modified system in turn determines the
spectral shape and energy content of the accelerated particles' 
distribution. The
creation of a shock precursor in the upstream region and the
amplification of the background magnetic field through streaming
instability are the main reasons to expect that the maximum achievable
momentum
may be appreciably different from the one calculated in linear theory
for the same value of the shock Mach number. The precursor, due to the
gradual slowing down of the upstream plasma would by itself be
expected to slightly decrease the maximum momentum: this effect can in
fact be seen in Fig.~\ref{fig:pmax}, where the dash-dotted line
shows the maximum momentum as a function of the Mach number when the
background field is not amplified. The reduced velocity in the
upstream plasma slows down the process of return to the shock for the
particles in the upstream section of the fluid. In the presence of
magnetic field amplification the maximum momentum is expected to
increase. This trend continues up to Mach numbers for which the
pressure in the form of accelerated particles cannot increase any more 
(saturation), so that the self-generated magnetic field also
saturates. On the other hand the compression factor (shock
modification) keeps increasing, so that the corresponding $p_{max}$ in 
Fig.~\ref{fig:pmax} shows a slightly decreasing trend (roughly the
same that one obtains in the presence of shock modification but
without field amplification, dash-dotted line). One should be careful
not to mistake this trend for a temporal trend. The behaviour of the
maximum momentum is to be interpreted only as a trend as a function of
the fluid Mach number at upstream infinity. In a more realistic
situation the Mach number can also change with time, which might
introduce additional effects, not taken into account here. On the
other hand, if the conclusions of \cite{lc83b} remain valid for this
more complex scenario, one should expect that most of the acceleration 
of cosmic rays occurs between the end of the free expansion phase and the
beginning of the Sedov phase. This is the phase we concentrated our
attention upon, so that the assumption of temporally constant Mach 
number adopted here is justified. 

The most important equation derived in this work is Eq. \ref{eq:tacc}
for the acceleration time in the presence of shock modification. It is
important to realize that although the calculations shown here have
been carried out by adopting the theory of \cite{amato1,amato2} for
the evaluation of the modification and field amplification, Eq. 
\ref{eq:tacc} can be applied in a very general way. In fact all the
quantities involved in the evaluation of Eq. \ref{eq:tacc} can be
determined using very different approaches, for instance Monte Carlo
simulations, while Eq. \ref{eq:tacc} can be adopted to estimate the
{\it instantaneous} value of $p_{max}$. Moreover, since
Eq. \ref{eq:tacc} directly contains the diffusion coefficients and the
dynamical quantities (e.g. fluid velocity), it can be applied
independently of the specific mechanism of wave production. If the
non-resonant wave generation mechanism recently proposed by 
\cite{bell2004} were to be responsible for the wave production and 
particle scattering,
this would change the mathematical form of the diffusion coefficient,
and possibly the dynamical reaction of the plasma (if the amplified
field reaches a dynamically important value), but Eq. \ref{eq:tacc}
could still be applied unchanged. 

The most relevant shortcoming of the calculation illustrated here is
in the fact that we used the assumption of quasi-stationarity in the
upstream fluid (not downstream). This assumption is fundamental to 
allow us to apply the method of Laplace transform, which can be used only for
linear differential equations. At the same time quasi-stationarity is
also required in order to use our previous
calculations (\cite{amato1,amato2}) for the shock modification and 
field amplification. These are actually two different
aspects of the same problem of this approach, which however cannot be
avoided if an analytical attempt has to be made to calculate the 
maximum momentum in these complex situations. 

It should however be stressed that the shortcomings just discussed also
apply to the approach of \cite{lc83a,lc83b}: their
calculations are only valid for linear shocks ($P_{CR}\ll \rho_0 u_0^2$)
so that the velocity and density profiles at the shock are spatially
constant and given. On
the other hand the diffusion coefficient, which is self-generated in
their approach, is time-dependent because the particle distribution
function is time-dependent. Despite this, the authors fix the
diffusion coefficient (and the distribution function of the
accelerated particles) and proceed by applying the method of the
Laplace transform. In this sense our approach should be considered as
the extension of the work of \cite{lc83a,lc83b} to the case in which
the dynamical reaction of the accelerated particles is not neglected. 
The common finding of several independent approaches to particle
acceleration at shocks is that shock modification manifests itself in
a prominent way, as a consequence of efficient acceleration, therefore 
our extension seems particularly important at this time. 

We conclude this section by providing a short, itemized summary of our
main scientific results:

1) the maximum momentum of accelerated particles can be as high as 
$\sim 10^6$ GeV/c for protons, when all effects are taken
into account. The approach of \cite{lc83a,lc83b} would give
$p_{max}\sim 6\times 10^6$ GeV/c when the magnetic field is assumed
spatially constant to its maximum value at the shock surface. The same
approach for $\delta B/B_0\approx 1$ would reproduce the well known
value $p_{max}=10^5$ GeV/c for $B_0=10\mu G$ and $u_0=5000\rm km/s$. 

2) the net effect of the shock modification is to create a precursor
   in both the fluid velocity and magnetic field strength. The former
   decreases while moving towards the shock while the latter
   increases. Both phenomena contribute to lowering $p_{max}$ with
   respect to the value $\sim 6\times 10^6$ GeV/c. 

3) Provided the magnetic field is amplified by the streaming
   instability of cosmic rays, the value of $p_{max}$ is weakly
   dependent upon the exact spectrum of the magnetic turbulence. 

4) The expression provided in Eq. \ref{eq:tacc} is the only expression 
presently available that allows one to calculate the acceleration time
for cosmic ray modified shock waves. This expression is very general 
and can be applied independently of the specific treatment of the
shock modification and reduces to the well known formula for weakly 
modified shocks.

\end{document}